\input{aipcheck}

\documentclass[
    ,final            % use final for the camera ready runs
  ]
  {aipproc}

\layoutstyle{6x9}

\usepackage{fabienne}
\begin{document}

\title{Recent Studies of Little Higgs Models in ATLAS}


\author{Fabienne Ledroit}{
  address={LPSC; 53 avenue des Martyrs; F-38026 Grenoble CEDEX}
}

\newcommand{\atlas}[0]
{%
{\sc Atlas}\xspace%
}

\begin{abstract}
The ATLAS Collaboration at the LHC continues investigating the possibility
to detect particles predicted by Little Higgs models.
In this talk, the latest results on the \Z/\W \ $h$ decays
and on the hadronic decays of the new gauge bosons \zh/\wh \ are reviewed.
\end{abstract}

\maketitle

%%%%%%%%%%%%%%%%%%%%%%%%%%%%%%%%%%%%%%%%%%%%
%% MAINMATTER
%%%%%%%%%%%%%%%%%%%%%%%%%%%%%%%%%%%%%%%%%%%%

\section{Introduction}

The Higgs sector of the Standard Model (SM) contains an instability 
due to the radiative corrections to the Higgs boson mass;
this is known as the hierarchy problem. Various theories and models
have been designed to solve this problem. One of them is called
the Little Higgs model and was proposed a few years ago \cite{ACG}. 
This is an effective model in which the Higgs
is a pseudo-Goldstone boson resulting from a spontaneously broken global
symmetry at a very high scale. Such a Higgs boson remains light, protected
by the approximate global symmetry. 
Here we concentrate on one of the several implementations
of the model, called the {\it Littlest Higgs} model \cite{littlest}.
This particular model includes a $SU(5)$ to $SO(5)$ breaking, 
and the following heavy new particles:
scalar particles $\phi^0$, $\phi^+$, $\phi^{++}$,
new 2/3 charge heavy quark (\T),
and heavy partners of the electroweak gauge bosons \ah,\ \zh \ and \whpm.

\section{Phenomenology at the LHC}
The Higgs boson (\h) properties remain those of the SM.
Production cross sections and branching ratios of the heavy new particles
have been computed in \cite{Burdman,Hanetal},
leading to the following search strategy:
\begin{itemize}
\item
The $\phi^{++}$ scalar is produced in $W^+W^-$ fusion. The dominant decay mode 
is $\phi^{++}\to W^+W^-$. The corresponding cross section is small and the SM background large, 
so this particle would be difficult to observe. Nevertheless, the prospects for its detection
have been studied and reported in \cite{Azuelosetal}.
\item
The heavy quark \T can be produced singly in the $bq\to Tq'$ channel via \W exchange.
The \T \ then decays to $tZ$ or $th$. The corresponding cross section is small as well,
although larger than the cross section for the QCD production of a $T\bar T$ pair.
How well the \T \ could be detected by ATLAS was studied in \cite{Azuelosetal}.
\item
The gauge bosons \zh \ and \wh \ are produced via $q\bar q^{(\prime )}$ annihilation
and the cross section is much higher than for the $\phi^{++}$ and \T. Once their mass is fixed
(they are approximatly degenerate),
the only parameter is $\theta$, the mixing angle between the $W$ triplets,
and the cross section is proportional to $(\cot\theta )^2$.
The production cross section for the \ah \ boson is more difficult to predict
since the couplings are not entirely fixed by the model. Nevertheless,
some results on its search are shown in~\cite{Azuelosetal}.
\end{itemize}

In the following, we concentrate on the experimental search of \zh \ and \wh \
using the ATLAS detector. The \texttt{PYTHIA} event generator\cite{pythia} was used
to generate signal and background events. The events were passed through the ATLAS
fast simulation\cite{atlfast} which provides a parameterized response of the detector
to jets, electrons, muons, isolated photons and missing transverse energy.
The integrated luminosity is assumed to be 300~fb$^{-1}$, which corresponds 
to data collected during three years of running at high luminosity. 
The event selection includes the ATLAS trigger criteria.

\section{Heavy gauge boson searches}

Once produced, the heavy \zh \ and \wh \ bosons can have two types of decays:
the fermionic channels, which are the same channels as their SM partners,
and the bosonic channels. Denoting \vh \ for both \zh \ and \wh \ (and \V \ for \Z \ and \W), 
these are $V_H\to V h$ and $Z_H\to W^+W^-$, $W_H\to Z W$.

Among the fermionic channels, $Z_H\to l^+l^-$ and $\W_H\to l\nu$ 
are clearly the discovery channels. It was shown in~\cite{Azuelosetal}
that \vh \ can be discovered at masses of up to 6~TeV if $\cot\theta$ is large.
On the other hand, the  $V_H\to q\bar q^{(\prime )}$ channel is expected to be more difficult,
because of the high level of background. It was studied however, and the results are reported below.

\begin{figure}
  \includegraphics[height=.3\textheight]{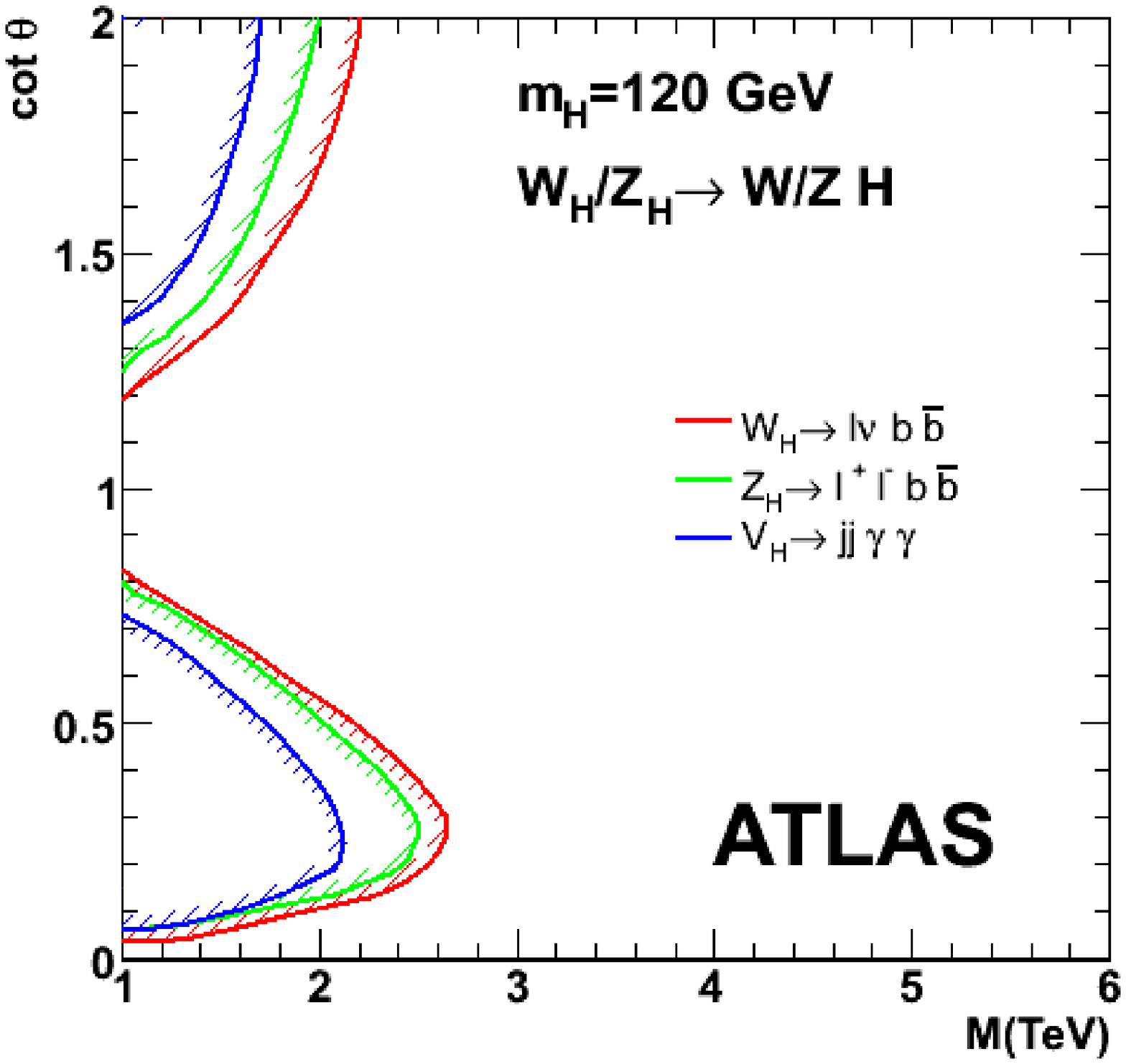}
  \includegraphics[height=.3\textheight]{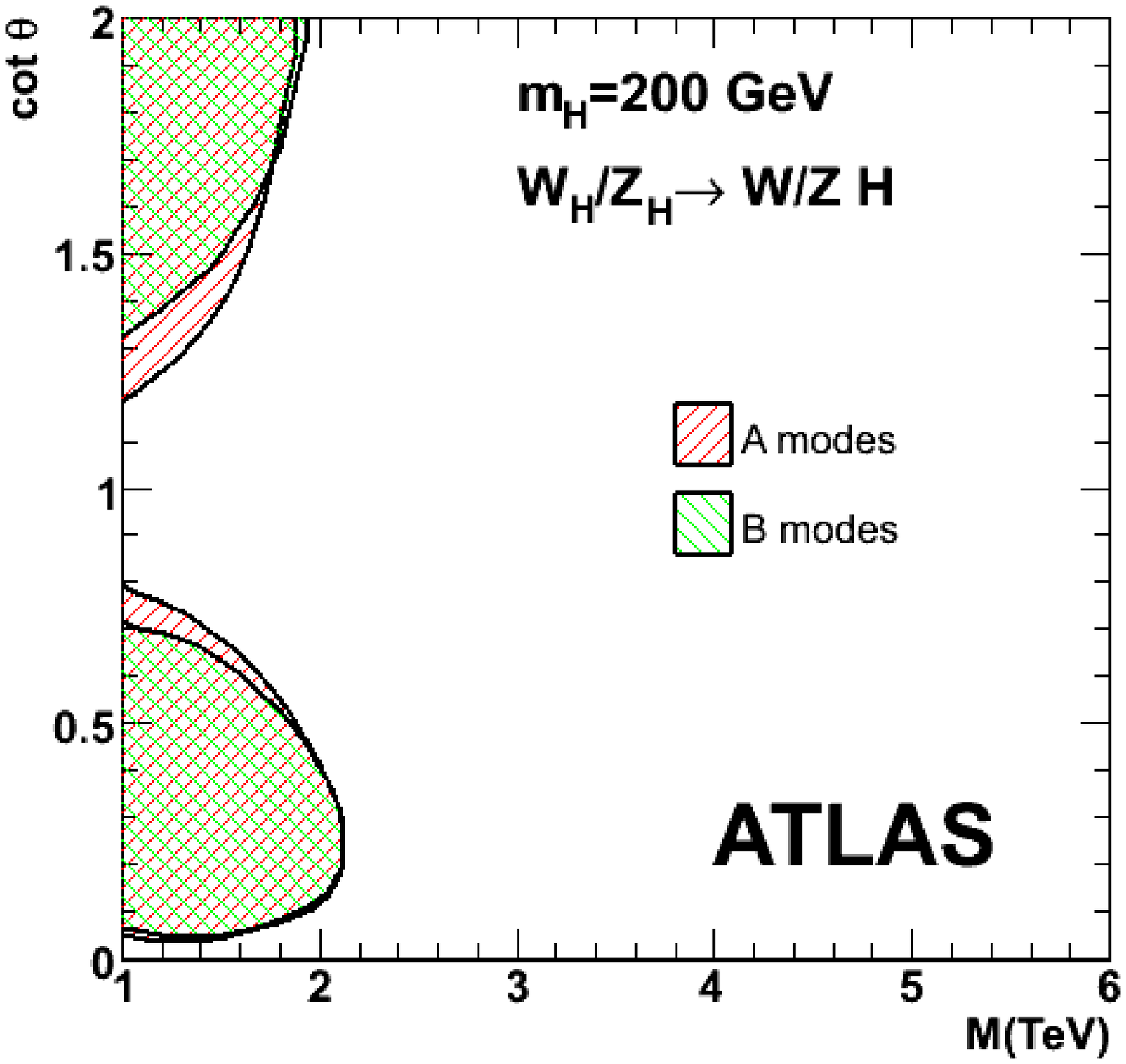}
  \caption{Discovery region with the $V_H\to V h$ channel (left: $m_h=120$~GeV, right: $m_h=200$~GeV) as a function of $M_{V_H}$ and $\cot\theta$.}
  \label{fig:mh120200}
\end{figure}

The observation of the $V_H\to V h$ decay is essential to test Little Higgs models.
Besides, it is dominant at low values of $\cot\theta$. A first study was performed
assuming that the Higgs boson is discovered with a mass $m_h=120$~GeV \cite{Azuelosetal}. 
The resulting exclusion contour in the $(M_{V_H},\cot \theta)$ plane is shown on Fig~\ref{fig:mh120200}~(left).
In the next subsection, we report on the results of a more recent study of this decay,
which assumes $m_h=200$~GeV.

\subsection{Search for the decays $V_H\to Vh$ assuming $m_h=200$~GeV.}

When $m_h=200$~GeV, the main Higgs decays are $h\to W^+W^-$ (73\%) and $h\to ZZ$ (26\%). Various $Zh$ and $Wh$ final states
have been chosen, as listed in Table~\ref{tab:mh200}, resulting from a compromise between cross section and signature~\cite{mh200+recap}.
\begin{table}
\begin{tabular}{cclll}
\hline
    \tablehead{1}{r}{b}{Mode}
  & \tablehead{1}{r}{b}{BR ($10^{-4}$)}
  & \tablehead{1}{c}{b}{Decay}
  & \tablehead{1}{l}{b}{Signature} 
  & \tablehead{1}{l}{b}{$\epsilon$} \\
\hline
A1 & 1.0 & $Z_H\to Zh\to l^+l^-W^+W^-\to l^+l^- l^+\nu l^-\nu$ & 4 leptons + $E_T^{miss}$ & 34\%\\
A2 & 3.0 & $W_H\to Wh\to l\nu-W^+W^-\to l\nu l^+\nu l^-\nu$    & 3 leptons + $E_T^{miss}$ & 12\%\\
A3 & 0.4 & $Z_H\to Zh\to jjZZ\to jjl^+l^- l^+l^-$              & 4 leptons + jets & 26\%\\
A4 & 0.4 & $W_H\to Wh\to jjZZ\to jjl^+l^- l^+l^-$              & 4 leptons + jets & \\
B1 & 6.8 & $Z_H\to Zh\to l^+l^-W^+W^-\to l^+l^- jjl\nu$        & 3 leptons + jets + $E_T^{miss}$ & 22\%\\
B2 & 0.8 & $Z_H\to Zh\to l^+l^-ZZ\to l^+l^- jjl^+l^-$          & 4 leptons + jets & 17\%\\
B3 & 2.4 & $W_H\to Wh\to l\nu ZZ\to l\nu jjl^+l^-$             & 3 leptons + jets & 15\%\\
\hline
\end{tabular}
\caption{\zh \ and \wh \ final states under study. The branching ratios are computed assuming $\cot\theta=0.5$,
the selection efficiencies $\epsilon$ assume a gauge boson mass of 1~TeV.}
\label{tab:mh200}
\end{table}
In all the modes, the main background is inclusive top pair production, where both tops decay semi-leptonically
and a third lepton can arise from a $b$ jet.
In the A1 and A2 modes, the missing transverse momentum is used to reconstruct the Higgs momentum,
assuming that the neutrino is colinear to the leptons. In addition, the \W \ mass constraint is applied
in the B1 and B3 mode. The A3 and A4 modes have indistinguishable final state.
The hadronic decay of high $p_T$ \W \ or \Z \ are reconstructed by looking for two high $p_T$ jets
with mass close to the \W/\Z \ mass or, if it fails, by taking the jet with the largest $p_T$.

The corresponding discovery reach is summarised on 
Fig~\ref{fig:mh120200}(right); it is similar to the $m_h=120$~GeV case, but the
bound on the \vh \ mass is weaker (6~TeV versus 2~TeV).

\subsection{Search for hadronic decays }

For $\cot\theta\approx 1$, BR($V_H\to Vh$) vanishes and the branching ratios to heavy quarks are\cite{Hanetal}:
BR($Z_H\to b\bar b$)=BR($Z_H\to t\bar t$)=1/8 and BR($W_H\to tb$)=1/4.
It was shown\cite{hadronic} that no convincing signal can be seen in the \zh \ case. However, the $W_H\to tb$
appears visible in the channel where the top decays to the $W(l\nu )b$ final state. 
One isolated lepton is searched for, and two $b$-jets tagged, one close to the lepton, 
one recoiling against the lepton. The neutrino 3-momentum is estimated from the reconstructed missing transverse 
momentum and assuming it is parallel to the lepton momentum.

\begin{figure}
  \hspace{-1cm}
  \includegraphics[height=.29\textheight]{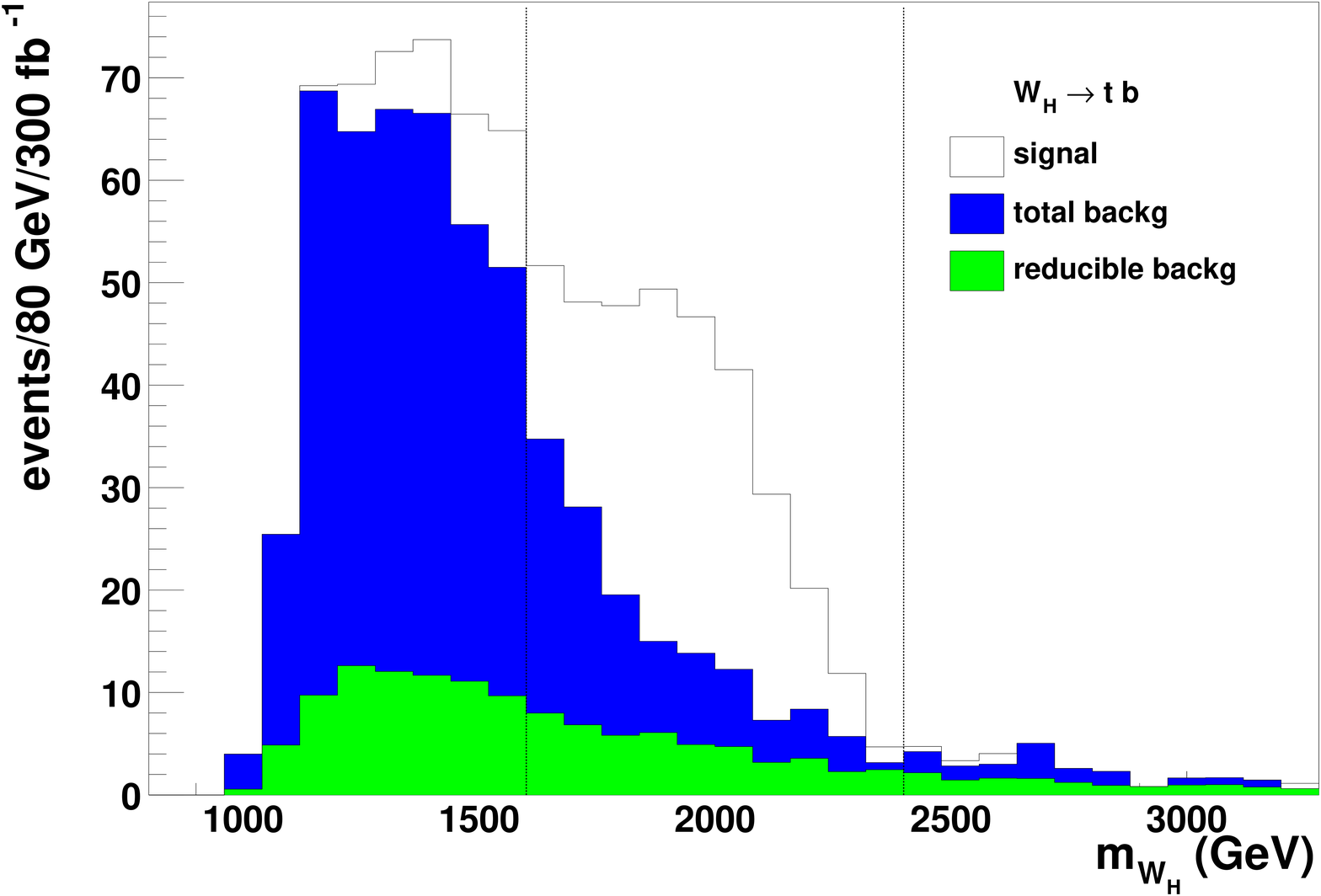}
  \includegraphics[height=.29\textheight]{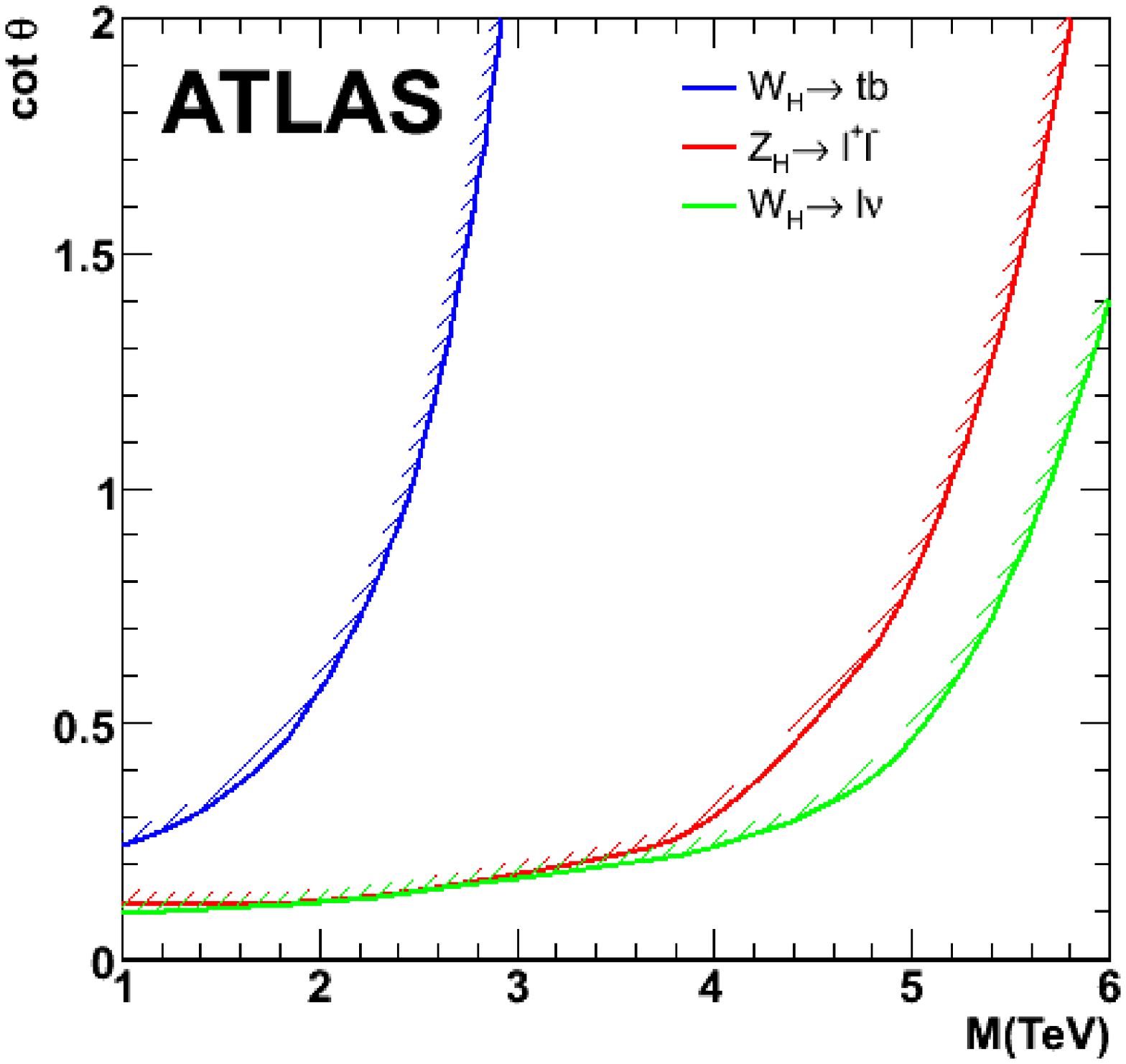}
  \caption{Left: reconstructed mass peak for $W_H\to tb$ ($M_{W_H}=2$~TeV). Right: discovery region with the fermionic channels as a function of $M_{V_H}$ and $\cot\theta$.}
  \label{fig:hadronic}
\end{figure}

The discovery reach is shown on Fig~\ref{fig:hadronic} in the same plane as before,
together with earlier results on the $Z_H\to l^+l^-$ and $W_H\to l\nu$ discovery channels~\cite{Azuelosetal}.
The $\cot\theta=1$ region which was missing in the $V_H\to Vh$ analyses 
is well covered up to $M_{W_H}=2.5$~TeV.

\section{Summary}
We have shown the latest examples of channels which can be used to test the predictions of Little Higgs models.
The discovery reach is found to be rather good, in particular for 
the new gauge bosons.
However, we are aware that the model family of the particular implementation studied here
is actually disfavoured by precision electroweak constraints. 
The introduction of a discrete symmetry called T-parity would allow to escape them.
Unfortunately, since it implies that the heavy partners can only be produced in pairs,
the phenomenology is drastically modified and would therefore require a new analysis.

%%%%%%%%%%%%%%%%%%%%%%%%%%%%%%%%%%%%%%%%%%%%%%%%
%% BACKMATTER
%%%%%%%%%%%%%%%%%%%%%%%%%%%%%%%%%%%%%%%%%%%%%%%%

\begin{theacknowledgments}
Thanks are due to the authors of these analyses:
J.E.~Garcia, S.~Gonz\'alez de la Hoz, M.~Lechowski, 
L.~March, E.~Ros and, last but not least, D.~Rousseau. 
\end{theacknowledgments}

%%%%%%%%%%%%%%%%%%%%%%%%%%%%%%%%%%%%%%%%%%%%%%%%
%% The bibliography can be prepared using the BibTeX program or
%% manually.
%%
%% The code below assumes that BibTeX is used.  If the bibliography is
%% produced without BibTeX comment out the following lines and see the
%% aipguide.pdf for further information.
%%
%%%%%%%%%%%%%%%%%%%%%%%%%%%%%%%%%%%%%%%%%%%%%%%%

%%%%%%%%%%%%%%%%%%%%%%%%%%%%%%%%%%%%%%%%%%%%%%%%
%% You may have to change the BibTeX style below, depending on your
%% setup or preferences.
%%
%%
%% For The AIP proceedings layouts use either
%%%%%%%%%%%%%%%%%%%%%%%%%%%%%%%%%%%%%%%%%%%%

\bibliographystyle{aipproc}   % if natbib is available
%\bibliographystyle{aipprocl} % if natbib is missing

%%%%%%%%%%%%%%%%%%%%%%%%%%%%%%%%%%%%%%%%%%%
%% You probably want to use your own bibtex database here
%%%%%%%%%%%%%%%%%%%%%%%%%%%%%%%%%%%%%%%%%%%
\bibliography{ledroit_fabienne}

\hyphenation{Post-Script Sprin-ger}
\begin{thebibliography}{8}
\expandafter\ifx\csname natexlab\endcsname\relax\def\natexlab#1{#1}\fi
\providecommand{\enquote}[1]{``#1''}
\expandafter\ifx\csname url\endcsname\relax
  \def\url#1{\texttt{#1}}\fi
\expandafter\ifx\csname urlprefix\endcsname\relax\def\urlprefix{URL }\fi
\providecommand{\eprint}[2][]{\url{#2}}

\bibitem[Arkani-Hamed et~al.(2001)]{ACG}
N.~Arkani-Hamed, A.~G. Cohen, and H.~Georgi, \emph{Phys. Lett. B} \textbf{513},
  232 (2001).

\bibitem[Arkani-Hamed et~al.(2002)]{littlest}
N.~Arkani-Hamed, A.~G. Cohen, E.~Katz, and A.~Nelson, \emph{JHEP}
  \textbf{0207}, 034 (2002).


\bibitem{Burdman}
G.~Burdman, M.~Perelstein and A.~Pierce,
\emph{Phys. Rev. Lett.} {\textbf 90}, 241802 (2003)
[Erratum-ibid. {\bf 92}, 049903 (2004)]

\bibitem[Han et~al.(2003)]{Hanetal}
T.~Han, H.~E. Logan, B.~McElrath, and L.~T. Wang, \emph{Phys. Rev. D}
  \textbf{67}, 095004 (2003).

\bibitem[Azuelos~et al.(2005)]{Azuelosetal}
G.~Azuelos~et al., \emph{Eur. Phys. J} \textbf{C39S2}, 13--24 (2005).

\bibitem[Sj\"ostrand~et al.(2001)]{pythia}
T.~Sj\"ostrand~et al., \emph{Comput. Phys. Commun.} \textbf{135}, 238 (2001).

\bibitem[Froidevaux et~al.(1998)]{atlfast}
D.~Froidevaux, E.~Richter-Wa\c{s}, and L.~Poggioli, {ATLFAST 2.0: a fast
  simulation package for ATLAS}, ATL-PHYS-98-131 (1998).

\bibitem[Ros and Rousseau(2006)]{mh200+recap}
E.~Ros, and D.~Rousseau, {Little Higgs studies in ATLAS}, 
  ATL-PHYS-CONF-2006-007 (2006).

\bibitem[Gonz\'alez de~la Hoz et~al.(2006)]{hadronic}
S.~Gonz\'alez de~la Hoz, L.~March, and E.~Ros, {Search for hadronic decays of
  $Z_{H}$ and $W_{H}$ in the Little Higgs model}, 
  ATL-PHYS-PUB-2006-003 (2006).

\end{thebibliography}

\end{document}